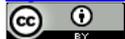

Scientific
Research
Publishing

# Commercial Technologies for Advanced Light Control in Smart Building Energy Management Systems: A Comparative Study

**Roufaida Laidi[a,b], Djamel Djenouri[a], Marc Ringel[c]**

(Affiliation): [a]*CERIST Research Center, Algiers, Algeria;* [b]*Ecole Nationale Suprieure dInformatique, Oued-Smar, Algiers, Algeria;* [c]*Nuertingen Geislingen University, Geislingen, Germany;*

Email: *rlaidi@cerist.dz; ddjenouri@acm.org ; marc.ringel@hfwu.de*

## Abstract

This work investigates the economic, social, and environmental impact of adopting different smart lighting architectures for home automation in two geographical and regulatory regions: Algiers, Algeria, and Stuttgart, Germany. Lighting consumes a considerable amount of energy, and devices for smart lighting solutions are among the most purchased smart home devices. As commercialized solutions come with variant features, we empirically evaluate through this study the impact of each one of the energy-related features and provide insights on those that have higher energy saving contribution. The study started by investigating the state-of-the-art of commercialized ICT-based light control solutions, which allowed the extraction of the energy-related features. Based on the outcomes of this study, we generated simulation scenarios and selected evaluations metrics to evaluate the impact of dimming, daylight harvesting, scheduling, and motion detection. The simulation study has been conducted using *EnergyPlus* simulation tool, which enables fine-grained realistic evaluation. The results show that adopting smart lighting technologies have a payback period of few years and that the use of these technologies has positive economic and societal impacts, as well as on the environment by considerably reducing gas emissions. However, this positive contribution is highly sensitive to the geographical location, energy prices, and the occupancy profile.

## 1. Introduction

The market of smart buildings is booming with the constant proliferation of new products and services. The competition in this domain between Information and communication technology (ICT) stockholders becomes aggressive, and companies constantly claim the potential of their products for energy savings.

While many trust the smart home technologies on providing sustainable energy management solutions, others are still doubting about the real benefits of these technologies [1]. The present study investigates (qualitatively and quantitatively) the use of smart technologies in buildings' energy management while focusing on lighting control systems and their contribution to energy saving. This study is not limited to the estimation of reduction in energy consumption due to the light control systems, but it also explores the impact of each feature offered by these technologies. Increasing the granularity of the study provides a deeper understanding of what may increase the energy-saving potential. The results of this study help the researchers, consumers, and manufacturers to improve the capability of the control systems in saving energy.

Contrary to most studies of existing literature, our comparative simulation study is not limited to energy measurement but investigates the problem from different facets including the economy (cost-benefit analysis), the environment (reducing emissions and sparing resources), and the social impacts (increase in the disposal income). We find these metrics more expressive and appealing to the readers than the amounts of saved energy. For instance "saving more than €10 000 over ten years, sparing the environment 2000kg of $CO_2$ yearly with a payback period of 6 months" drives a better understanding of the impact of such solutions than the value of saving "5000 kWh".

This study considers two cities with two different weather characteristics and regulatory environments regarding energy prices that impact energy usage: Algiers vs. Stuttgart. Algiers has more sunny days (notably in winter), which increases the potential of daylight harvesting. Germany has higher energy prices, which raises the economic benefits of energy saving in Stuttgart. The focus of the study is on light control, which is the second most energy consuming system in today's buildings after HVAC (heating ventilation and air conditioning). Moreover, this system has a visible impact on the comfort and productivity of buildings occupants. Smart bulbs are one of the low-cost easy-to-deploy technologies and one of the fastest growing products in the internet of things (IoT) markets. It is expected a constant growth over the next years, with penetration in tens of millions of households by 2020[2]. A survey of nearly 300 ICT devices adopted by consumers shows that smart lighting solutions are amongst the most demanded, which motivates the choice for this study.

Light control products in the market are explored. We accordingly extract the features contributing most in saving energy from those improving comfort and security. These features are daylight harvesting, dimming, scheduling, and motion detection. We translated them into simulation scenarios of a family house using *EnergyPlus* simulation tool where we consider two types of families.

The remainder of this paper is as follows, Sec.2 describes the related work. Sec.3 summarizes the ICT based light control devices that are available in the market. Sec.4 describes the simulation setup, while Sec.5 shows the simulation results. Finally, Sec.6 concludes the work.

## 2. Related Work

Due to the importance of energy management in buildings and the potential, it presents for energy saving, many research and development efforts have been devoted to optimizing energy usage in smart buildings. Researchers and engineers proposed control strategies that create environments able to satisfy their occupants' comfort needs while minimizing energy consumption. Due to their high energy footprint and their high impact on comfort and productivity, lighting and thermal systems are the most considered. Chew et al. [3] provided a review of solutions proposed for the optimization of lighting. They described the most common methods to reduce energy use, such as detecting the occupancy of rooms and daylight harvesting. They measured the potential of different solutions by the percentage of energy savings. The paper also discusses solutions trying to optimize the non-visual effects of light, such as reducing the human stress levels and increasing health and work performance. In addition to energy and lighting output optimization, technical aspects such as the connectivity between different components of the system and Visible Light Communication (VLC) have been discussed.

Home energy saving solutions are typically occupancy-detection-based, which motivates researchers proposing solutions focusing on the accurate detection of room occupancy. The occupancy information varies from binary (occupant vs. vacant) detection to a head-counting of people in real time. [4] considered occupancy-based lighting control in open-plan office spaces. Because of the importance of occupancy information, other researchers focused on the motion sensors deployment in buildings. [5] proposed to dedicate a pre-deployment learning phase to detect the most used spots in a building and next concentrated the motion sensors deployment in these spots. [6,7] considered the gaps in Passive Infrared (PIR) motion sensors' field of view (called sensing-holes). Inside these sensing holes, the sensor is incapable of detecting occurring motions, which leads to false negatives. The authors proposed a solution to overlap sensing areas in a way to cover these gaps and increase the accuracy.

Despite the rich state-of-the-art in smart building solutions, this domain is far from reaching maturity. There is a gap between current research and the existing commercialized systems. This gap makes the promises made by the state-of-the-art approaches in terms of reducing consumed energy different from the reality in gain provided by the commercialized solutions. Commercial solutions emphasis on aesthetics and remote control by including mobile application interfaces to allow users to adjust light output parameters and rely on this information and cloud data as input for their control algorithms. [3] addresses the different features proposed by lighting commercialized solutions and emphasizes this difference compared to the existing literature. A work that is wholly dedicated to commercialized solutions is [8], where the authors collected a database of over 300 devices ([9]) and grouped them into categories. They specified for each category the main features of the smart home devices and described how these features might reduce energy consumption. A detailed report of this study is available in [10]. Despite that this study analyzed many devices, it is limited to qualitative comparison, and the authors did not conduct any numerical study to estimate the impact of different features on energy use. Similarly, to the previous work, [11] is a technical report grouping smart devices and studying their features. The report includes some statistics about energy use using specific devices, but the authors relied on manufacturers claims to provide this information.

[12] studied smart buildings from customer perception by asking 29 participants about their opinions after living in a fully equipped smart house. The study included variable energy tariffs, smart metering, smart appliances, and home automation. The results show that the main incentives for adoption include monetary savings and the environmental footprint. On the other side, the high expenses, data privacy, and technical complexity are the main reasons for refusal. [13] shares with the present work the vision of the need for conducting numerical analysis on the benefits of new technologies. The authors evaluated the impact of replacing traditional lighting with LEDs in a campus university. However, the scope was limited to the effect of replacing a lamp with a more efficient one (CFL vs. LED lamps), without considering the features provided by the smart connected lamps, which are considered in our work. [14] and [15] provided frameworks to simulates the cost and benefit of smart home devices. The frameworks have some practical usefulness, although the focus of the two studies was on the impact of feedback-based devices (load monitors and home displays) on energy management. The impact of feedback has widely been studied, but the market is currently witnessing a decreasing interest to these" feedback-only" devices [1].

## 3. Commercialized Lighting Systems

This section is dedicated to describing commercialized solutions for modern lighting systems. The devices cited in this section are either solutions dedicated for lighting systems, like smart bulbs and switches or designed for more general purposes but can be used in smart lighting systems. Examples of these devices are smart hubs and sensors. In our presentation for each category, we start by enumerating the energy related and non-energy related features. We discuss next the used communication protocols and price ranges while giving a non-exhaustive list of products in the category.

### 3.1. Smart Light Control Devices

This category includes smart bulbs and switched, which have the same general goal as traditional bulbs and switches but with additional features explained in the next two sections.

#### 3.1.1. Smart Bulbs

Compared to traditional bulbs, smart bulbs offer a wider range of choices for control and interactivity. These features serve both comfort enhancement and energy saving. From the energy-saving perspective, smart bulbs allow scheduled timers to set up the wake, sleep, and away automation. Some bulbs use GPS to detect the user's smartphone exact location and automatically switch the lights on or off when the user reaches a certain distance from his house. Depending on the compatibility, some bulbs can also be combined with other smart home devices such as security cameras, thermostats, which allows them to benefit from the available information in these devices and optimize the light control, e.g., the weather information or the motion sensors available in smart thermostats. If a lamp is" If *This Then That" (IFTTT)*[16] compatible, it enables "recipes" creation to make light respond to certain triggers, such as changes in weather or motion detection. IFTTT is a website and app that lets the user link his devices in the cloud with what the website calls" recipes."

The mains comfort-related features offered by smart bulbs are their color-changing and remote-control capabilities. Color impacts the home's atmosphere: high color temperatures are suitable for office lighting and helps in staying awake, focusing, and working while low color temperatures generate a warm glow perfect for relaxing. Some lamps even can synchronize with certain movies and TV shows. However, the color-changing feature is not available in all smart lamps but only in high price ones. Another comfort increasing feature is the remote-control option from a mobile device which allows the user to adjust the brightness, the color, or both from any location, even outside the house.

Smart lamps' efficiency increases when being part of a network of sensors and other smart devices. Information provided by other elements in the network allows them to be aware of real-time changes in the environment and consequently adapt their output to these changes. Some bulbs create their mesh network requiring the existence of a hub to communicate with other devices and the user's smartphone. This is mainly due to the most common wireless lighting protocol, *ZigBee* [17]. Philips Hue LEDs [18], Osram Lightify LEDs[19] , Belkin WeMo LEDs[20] are examples of smart bulbs in the market that communicate with *ZigBee*. Because smartphones are generally not enabled with ZigBee, a hub bridging between the devices is required. On the other side, lamps like Lifx A19[21] can directly connect to the WiFi network. Some manufacturers opted for *Bluetooth* to pair their lamps directly with the mobile device, like those produced by GE [22]. The lamps do not require a bridge or to connect to the home network, but they cannot be monitored outside the Bluetooth range, e.g., outside the house. Using a wireless speaker like Amazon Echo or Google Home as a hub allow vocal light control. However, this feature is also offered by some bulb's mobile application.

Depending on the offered features, the price per bulb ranges from €15[1] mark up to €70. Most available lamps are A-shape bulbs with standard-size E-shape screw-in bases; which limits their fitting compatibly to other types of fixtures, e.g., candelabrum.

### 3.1.2. Smart Switches

Smart wall switches allow automating the on/off wall switches instead of automating the bulb. It is useful to replace a single switch wiring multiple bulbs or to automate small-size screw-in bases bulbs for which smart bulbs are not available. Smart switches have similar energy savings features as those of smart bulbs, by proposing timers, smart dimming, and the possibility to optimize lighting's efficacy when combined with sensors.

More sophisticated, and expensive, smart switches offer full-color touch screen displays, built-in speakers and microphones for intercom functionality, e.g., the Wink Relay[23]. However, these switches are not achieving the success desired by their manufacturers due to the high prices they come with while proposing similar or limited options compared to smart bulbs.

Similarly to connected bulbs, the majority of smart switches connect either through WiFi (e.g., Belkin WeMo switch[24]) or ZigBee (e.g., GE wall-in smart switch [25]). Switches communicating by ZigBee requires a hub to connect with the smartphone. Lutron[26], a company specialized in smart lighting, uses its own proprietary Lutron signal.

Smart switches cost around €50, which makes them a pricier choice compared to smart bulbs. Scaling up the house lighting system with smart switches is not as cost efficient as with smart bulbs unless they are used to control multiple bulbs. Using smart switches may result in less energy saving compared to bulbs even by applying the same control strategy in case the used lamps controlled by the switches are not efficient, e.g., using a smart switch to control a $60W$ incandescent lamp. On the other side, smart bulbs are all LED energy efficient lamps with wattage less than $11W$. For all these reasons, we consider smart bulbs in our simulation study.

## 3.2. Smart Hub and Wireless Speakers

Bluetooth LE, Lutron Clear Connect, WiFi, Z-Wave, and ZigBee are communication protocols included in commercialized home automation solution. The role of a home automation hub is to unify the connected devices and making them talk to each other. Hubs also facilitate the control of these devices by grouping them all in a smartly designed, attractive, and intuitive application instead of using a dedicated interface for each device. The price of a smart hub starts from €49 and may reach up to €350 for those supporting many standards and endowed with higher security, displaying interfaces and extended hardware. The support for IFTTT (If This Then That) is also useful, as it offers different ways to configure and trigger the connected devices. Wireless speakers such as Google home[27] and Amazon alexa[28] allow the user to control their compatible devices by voice commands.

## 3.4. Real Time Detection Using Sensors: Motion and Light Sensor

---

1 Prices were established based on a search in the commonly used internet marketplaces such as amazon or Alibaba.

A home is smarter if it is capable of adjusting the controlled components (such as lighting system) depending on the dynamic of its environment. The sensors are the eyes that allow to "see" within the building and respond to changes. The market offers a full band of sensors for smart homes with goals different goals ranging from energy, comfort, and security. We may cite water leak and freeze detectors from Roost[29] and Honeywell[30], Ecobee's Room temperature, and occupancy sensors[31], and weather sensors[32]. However, in this section, we focus on those affecting the performance of the smart light: occupancy and light sensors. In smart homes, doors and windows opening sensors are popular. Open/close sensors can be attached to a door or window and alert the user should it detect either a window or a door being opened. These sensors are used for security, but also have applications in light control and may be used to turn the lights on in case of user's entrance and off after a while with no motion.

The most common motion sensors are the PIR (passive infrared) sensor. They can detect bodies that physical emit energy (heat) as they cross their sensing ranges. Other types of motion sensors include micro and ultrasonic waves, and sometimes a combination of multiple types. Sensors may be purchased individually or as part of a bundle, e.g., Ring[33] and Samsung Smartthings motion sensor[34]. Most of the motion sensors we reviewed use the Z-Wave protocol to communicate. Motion coverage and sensitivity are vital factors for the efficiency of sensors. Some motion sensors, e.g., the Everspring Compact Z-Wave Motion Sensor[34], have an adjustable sensitivity. Having an adjustable range can be hugely beneficial, depending on the monitored space. Some also come with a feature called 'Pet immunity,' which do not detect a movement from being under a noted maximum size. For example, the Ecolink Z-Wave Plus Motion Detector[36] is immune to movement from pets less than $25kg$. The user must be accurate in placing its sensors to avoid false positive and negative alerts (a subject studied in the literature, e.g., [5,6,7]).

Finally, some motion sensors are meant for specific tasks. For instance, the Philips Hue motion sensor[37] is uniquely designed to activate lights when a movement is detected. Light sensors can turn off and on the light according to the indoor ambient light level, but they are less commonly used and less available in the market. Table 1 summarizes the smart home appliances related to the control of lighting system (they are also used in controlling thermal systems), their proposed features, and how these features may contribute in energy saving.

| Smart home energy technology | | Features& Functionality | Energy impacts |
|---|---|---|---|
| Sensors | Motion sensors | -Real time detection of occupied areas | - Consume energy only in occupied locations |
| | Light sensors | -Real time detection of light level | -Enable daylight harvesting<br>-Enable dimming in function of daylight |
| Light actuators | Bulbs | -Scheduling, adapt with occupancy and/or daylight<br>-Remotely controllable<br>-Light level & color control | -Replacement of traditional bulb with LEDs<br>-Smart and rules based control<br>-Dimming, daylight harvesting. |
| | Switches | -Scheduling, adapt with occupancy and/or daylight | -Smarter management of bulbs<br>-Feedbacks |

| | | -Remotely controllable | |
| | | -Possibility of controlling multiple bulbs | -Dimming, daylight harvesting. |
| Hubs | | - Enable/manage interactions between devices | - Trigger-based control in function of real-time data received from other sources |
| | | - Remotely control multiple devices | -Scheduling feature across connected devices |
| | | - Talk different protocols | |
| | | - Rule based control | |
| | | - Different ways of interactions (eg. vocal) | |

Table 1: Summary of smart home appliances features and their energy saving potential

## 4. Simulation Study

In this section, the gain from adopting smart lighting is investigated by simulation. Without loss of generality, we select a single-family house environment for this study. The conclusions, though, can be generalized to any residential buildings. To define simulation scenarios, we first enumerated different features with a positive impact in reducing energy consumption and then grouped these features into categories and created a combination of scenarios from these features. We also present some evaluation metrics to measure the economic, environmental, and social impact of embracing smart lighting. Finally, we selected the simulation environment and materials. All this is explained in the following.

### 4.1. Simulation Scenarios

Based on insights from Sec.3, we selected features of smart lighting products that impact the consumed energy. These features may be divided into two categories, 1) occupant-unrelated features and, 2) occupant related features. Eq.(1) explains the impact of the two categories and presents the consumed energy as the multiplication of the lamp's power, $P$, and time of operation, $T$.

$$E(Wh) = P(W) \times T(h) = V(V) \times I(A) \times T(h). \quad (1)$$

, where $V$ is voltage in Volts, and $I$ is current in Amperes. The first category groups the features that reduce the consumed energy by reducing $P$ or $T$ independently from the user's behavior. In this category, we include dimming and daylight harvesting. The energy consumption in case of presence of the dimming is presented in Eq.2

$$E = a \times V \times I \times T, \quad (2)$$

where $a \in [0,1]$ is the duty cycling rate for cutting off power to the light bulb i.e., the power will be turned *On* and *Off* multiple times per second depending on $a$. The human eyes cannot notice the flickering and detect a constant flow of light. In case of light control, $a$ depends on the available daylight, the more daylight available, the more $a$ drops until achieving the value of 0 (when the available daylight satisfies the required lighting level). Furthermore, daylight harvesting may operate without dimming; That is, by reducing the operation time of the lamp and limiting it to periods of absence of daylight, i.e., between dusk and dawn or in case of cloudy weather. Because this category describes the implication of daylight harvesting in control, we call it daylight harvesting category, which results in three models: {no-daylight harvesting, daylight harvesting, daylight harvesting + dimming}.

The second category is the occupants' related category, which aims at relating the time of operation of the bulb to occupant presence inside the house.

In all the simulation scenarios, the family house is supposed occupied by a family of four members, and two occupancy profiles are considered that can serve as stylized profiles to model common patterns for both Central Europe and Northern Africa:

*Profile 1*: Continuous random occupancy along the day, while weekdays and weekends are similar. For instance, this profile may correspond to families with retired elderly persons, under school-age children, or a home staying adult.

*Profile 2*: Predictable long periods of non-occupancy in the middle of weekdays that correspond to regular working hours. Weekends are similar to *Profile 1*. This profile reflects families with working adults in regular working hours, and school enrolled children.

Table 2 shows the two methods used in this study to measure the time of operation of both profiles, i.e., scheduling vs. motion detection. Note that during sleeping periods (at night) where people are sleeping and do not make any movement, mention sensors report non-presence. These periods are then equivalent to non-occupancy (vacant) and noted such for the sake of simplicity.

In both methods, the goal is to define occupancy and vacancy periods, and the system will be running in occupation periods and turned-off in vacancy periods. In the case of "Scheduling", a fixed time is set by the user to turn the system on or off. In the case of "Motion *Detection*", the occupancy information is provided by a motion sensor included in the system. The addition of a motion sensor allows detecting the dynamics in occupant's behavior compared to the scheduling method that supposes the occupants' arrival and departure times are fixed and known a priori. We simulate the events detected by the sensors as percentages that vary according to the time of the day. In the simulation, the occupancy is presented as percentages (probability) during periods. We also consider one month of holidays where the family leaves the house 15 days in winter and 15 days in summer.

| | Scheduling | | Motion Detection | |
|---|---|---|---|---|
| | Time | House State | Time | % of occupancy |
| *Profile1* | 6 am to10 pm | Occupied | 6 am to7 am | 50 |
| | | | 7 am to8 am | 75 |
| | | | 8 am to 9:30 pm | 100 |
| | | | 9:30 pm to10 pm | 75 |
| | 10 pm to 6 am | Vacant | 10 pm to 6 am | 0 |
| *Profile2* | 6 am to 8 am | Occupied | 6 am to7 am | 50 |
| | | | 7 am to 7:30 am | 75 |
| | | | 7:30　am to 8 am | 100 |
| | 8 am to 6 pm | Vacant | 8 am to 6 pm | 0 |
| | 6 pm to10 pm | Occupied | 6 pm to 6:30 pm | 50 |
| | | | 6:30 pm to 7 pm | 75 |
| | | | 7 pm to 9:30 pm | 100 |
| | | | 9:30 pm to10 pm | 75 |
| | 10 pm to 6 am | Vacant | 10 pm to 6 am | 0 |

Table 2: Time of operation: scheduling vs. motion detection estimation for *Profile 1* and 2

| | | No Daylight harvesting | Daylight harvesting | Daylight harvesting + Dimming |
|---|---|---|---|---|
| **No Profile** | | Baseline — 0 € | DH — 399.35€ | DH+Dim — 399.35€ |
| **Profile 1** | Preset Schedule | Sched 1st — 139.93€ | Sched 1st+DH — 399.35€ | Sched 1st+DH+Dim — 399.35€ |
| | Motion Dtection | MD 1st — 374.07€ | MD 1st+DH — 399.35€ | MD 1st+DH+Dim — 399.35€ |
| **Profile 2** | Preset Schedule | sched 2nd — 139.93€ | Sched 2nd+DH — 399.35€ | Sched 2nd+DH+Dim — 399.35€ |
| | Motion Dtection | MD 2nd — 374.07€ | DM 2nd +DH — 399.35€ | MD 2nd +DH+Dim — 399.35€ |

Connected bulb — 19.99€ | Light/motion sensor — 24.90 € | Motion sensor — 21.29 € | 1 per room | Hub — 84.14 € | 1 per house

Fig.1 presents the 15 resulting simulation scenarios for the lighting system. They resulted from combining the two categories of features, daylight harvesting, and occupancy related, with the two profiles. The figure also shows the devices included in each scenario and the resulted cost.

## 4.2. Evaluation Metrics

We technically study the effects of installing smart lighting control systems while considering economic, environmental, and social aspects. The metrics used for each category are explained hereafter.

### 4.2.1. Technical and Economical Metrics

***Total Consumed Energy:.*** This metric shows the energy (in *kWh*) consumed by the house during one year for different simulation scenarios. This value is provided in our case by a simulation tool, as explained in Sec.4.3.

***Total Energy Cost:.*** The economic cost of one year of consumed energy taking into account the different regulatory frameworks. The price unit of €0.3048 per kWh is taken for Germany[38]. In Algeria, the government charges the first 125 KWh of consumed electricity for €0.014 per kWh, and €0.033 per kWh for the rest[39]. The low prices of energy in Algeria are due to the subsidy policy of electricity provided by the government.

***Payback Period:.*** It is the amount of time required for inflows coming from reducing energy consumption to meet the initial cost of the installation of the home control system. Eq.3 shows the formula of the payback period.

$$PB = \frac{Initial\ investment}{Annual\ inflows} \quad (3).$$

***Net Present Value (NPV):.*** It is the cash equivalent now of a sum receivable at a later date. If the money is not spent on installing the devices and is banked; instead, the result includes both the initial sum and the interest earned. NPV is a technique where cash inflows expected in future years are discounted back to their present value. It is calculated using a discount rate equivalent to the interest that would have been received on the sums if the inflows had been saved. NPV is given by Eq.4.

$$NPV = \sum_{n=0}^{n} \frac{C_n}{(1+r)^n} - initial\ investment, \quad (4)$$

where $C_n$ is the inflow at year $n$, and $r$ is the discount rate. We set $n = 10$ years, and $r = 5\%$. This mirrors the lifetime of equipment as guaranteed by most producers and the average interest rate, including inflation. A positive NPV means that the project is worthwhile.

***Internal Rate of Return (IRR):.*** The IRR is the rate of interest that reduces the NPV to zero. More detailed explanations of these concepts can be found in [40].

Figure 1: Simulation scenarios for lighting system

### 4.2.2. Environmental Metric: Gas Emissions

The impact of reducing energy consumption using smart devices on the environment is investigated by measuring the gas emissions due to power use. This is the amount of gases the house generates during one year of energy consumption. The considered gases are: Carbon dioxide $CO_2 (kg)$, Nitrogen dioxide $NO_2 (g)$, Sulfur dioxide $SO_2 (g)$, Carbon monoxide $CO$ $(g)$, and Methane $CH_4 (g)$. One $KWh$ generates respectively $0.516 kg$, $0.44g$ ,$0.290g$, $0.230g$, and $0.184g$ of these gases[41].

### 4.2.3. Social Metric: Additional Disposal Income

Disposal income is the amount of money that households have available for spending and saving. Through energy savings, household incomes will increase due to lower energy bills. It is depicted in the discounted annual inflows, measured using Eq.5,

$$ADI = \sum_{n=0}^{n} \frac{C_n}{(1+r)^n} \,.(5)$$

### 4.3. Simulation Environment and Materials

The house considered in our simulation is composed of two flours and seven main pieces: two bedrooms, living room, kitchen-diner, bathroom, toilet, and hall. The house template is provided by *DesignBuilder* [42] that enables the $3D$ modeling and energy simulation using *EnergyPlus* [43]. The later is the $2^n d$ most used simulator by researchers after Matlab [44]. The simulator provides the weather data for each of the two cities considered in the simulation (Algiers and Stuttgart).

## 5. Simulation Results

The obtained results of the comparison metrics in all scenarios are presented in this section. The following abbreviations are used:

**DH:** Daylight Harvesting **Dim:** Dimming **Sched:** Scheduling   **MD:** Motion detection

### 5.1. Technical and Economical Metrics

Fig.2 presents simulation results for energy consumption for each scenario. Daylight harvesting (DH) and dimming (DH+Dim) reduce the energy consumption by almost 11%, with a slight decrease in energy consumption for Algiers. This decrease is due to the longer sunshine hours the region is exposed to (2847 hours/year for Algiers vs. 1662 hours/year for Stuttgart)[45]. The savings increase when including management og the time of operation (predefined schedule or dynamic occupancy detection with motion sensor). A scheduling strategy (sched+) spares the third of energy usage for *Profile 1* and the two thirds for *Profile 2*. Savings are remarkable for *Profile 2* because of the long periods of occupants absence, i.e., long periods where lighting is not required. Including a motion sensor (MD+) leads to better results because of the accurate tracking of the presence. However, it requires additional hardware with a double cost for installation. Daylight harvesting and dimming are less effective for *Profile 2* since the inhabitants are mainly absent during day hours.

The results show that the occupancy pattern is the main factor in designing a smart lighting architecture. The inclusion of motion detection sensors is more profitable for families with dynamic occupancy habits during daylight hours. Scheduling is a suitable, budget-friendly, and an easy to deploy solution for residents with regular habits. These conclusions should also push the manufacturers to provide further options for schedule setting and occupation detection in their products.

Fig.3 presents the energy cost for one year. Because the government in Algeria partly subsidize the energy price (which makes it cheaper compared to Stuttgart) the vertical axis shows a large discrepancy in energy cost between the two regions. This difference may reduce the importance of some features for customers, e.g., daylight harvesting and dimming reduces less than €30 per year for an Algerian *Profile 2* family. It may push the customers toward low-cost scheduling installations. The same scenario saves almost €200 in Germany, which covers the initial cost of the additional hardware within the first year as shows the payback periods in Fig.4.

This also impacts the Net Present Value (NPV) (Fig.5) and the Internal Rate of Return (IRR) percentage (Fig.6). The scheduling-only scenarios show a small payback period and a high IRR, as the initial cost of this scenario includes only the cost of purchasing the lamps (scheduling being a feature in smart lamps), and it does not require any other sensors or hub. However, the NPV highly encourages to consider extended installations, especially in Germany, where the gain over ten years may reach €5500 .

## 5.2. Environmental Metric: Gas Emissions

Fig 7 shows the ecological footprint for the lighting scenarios. The two cities have very close results for the environmental metric with a slight decrease in Algeria when light harvesting in included. The impact of these solutions on reducing energy consumption translates to a positive impact on the environment. $CO_2$ emission may be reduced to the half for *Profile 1* and the quarter for *Profile 2*. Manufacturers can display similar statistics of environmental lighting footprint through a web or a mobile platform to increase individual's awareness and engagement.

## 5.3. Social Metric: Additional disposal income

Fig.8 shows the average inflows for ten years. The results confirm that smart lighting technologies do not only pay for themselves (as shown in Fig.4 for the payback period) but also result in more saving for the family in the long term. While the payback period encourages embracing low-cost solutions, the ADI indicator pushed toward the extended installation. Since the expected lifetime of such devices may exceed ten years, prioritizing such long-term benefits is more reliable

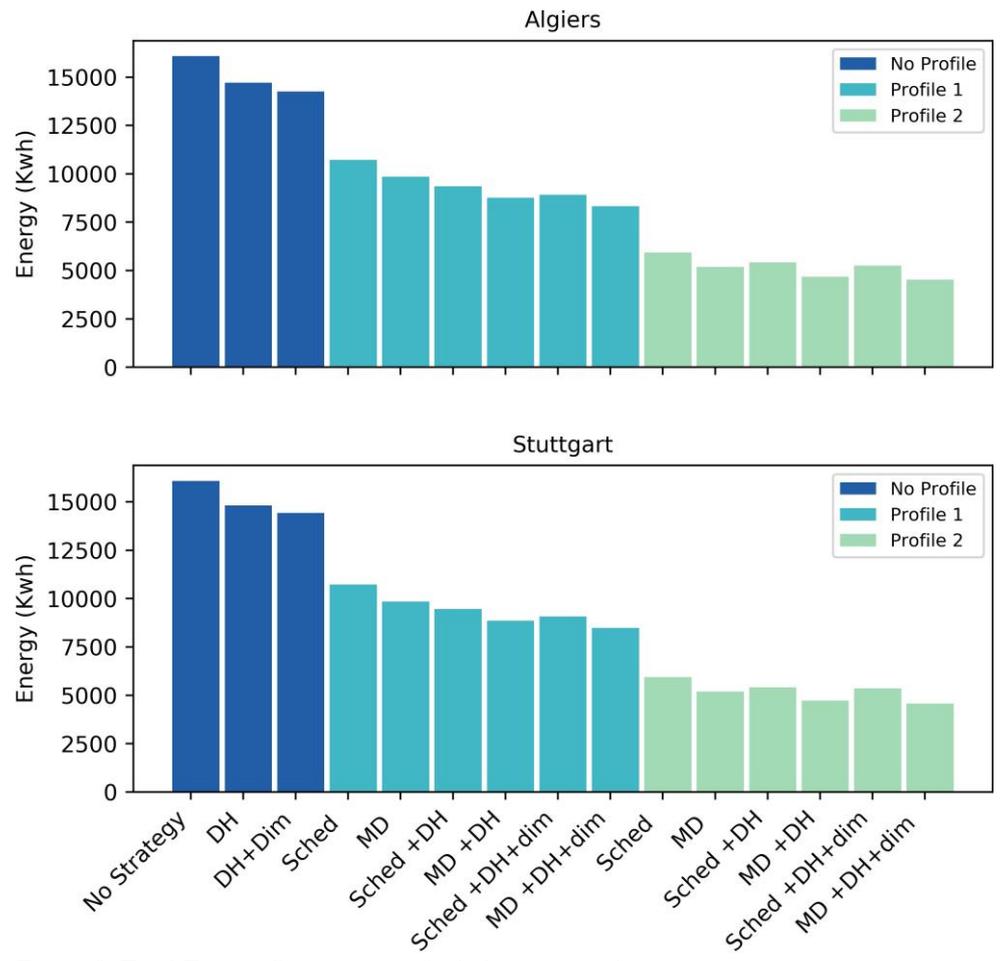

Figure 2: Total Energy Consumption for lighting control strategies

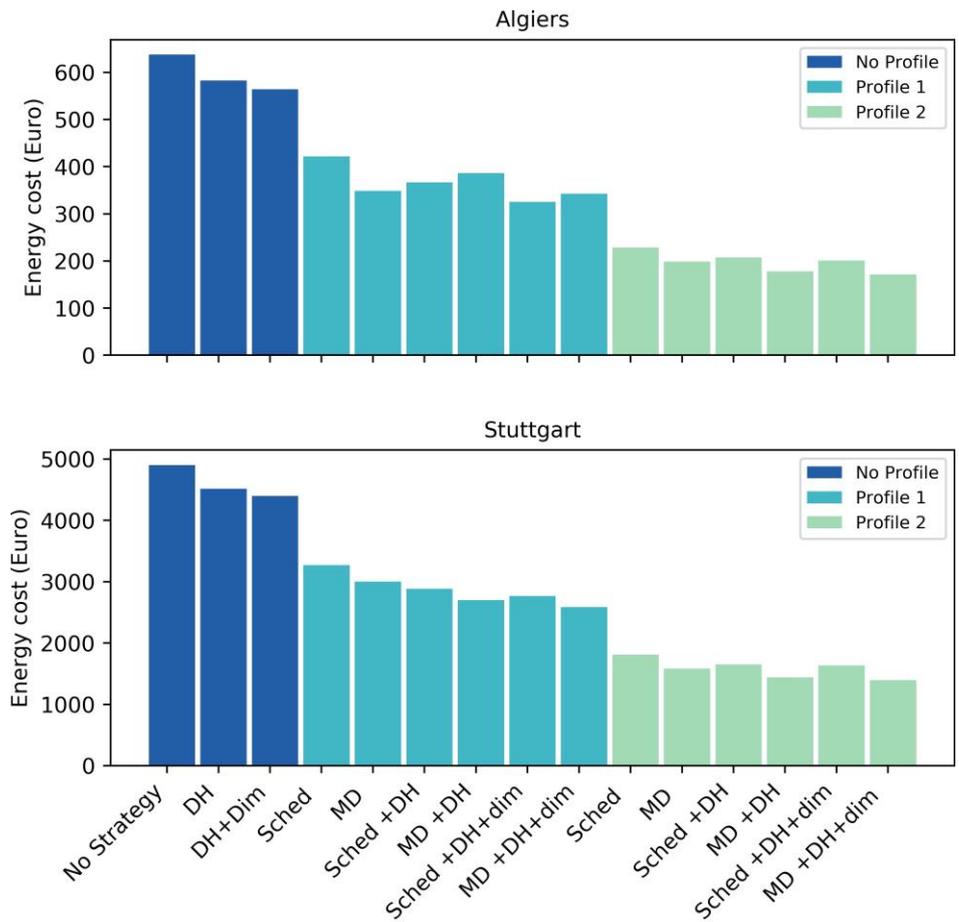

Figure 3: Energy cost for lighting control strategies (€)

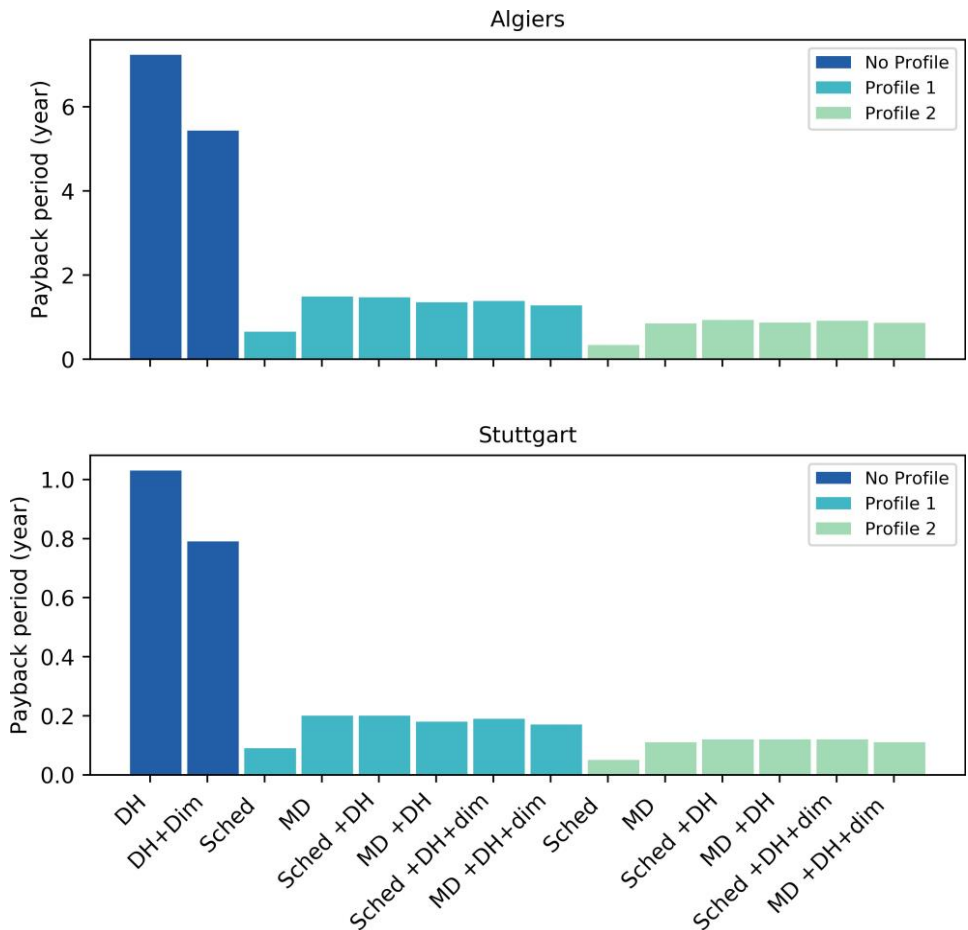

Figure 4: Payback period for lighting control strategies

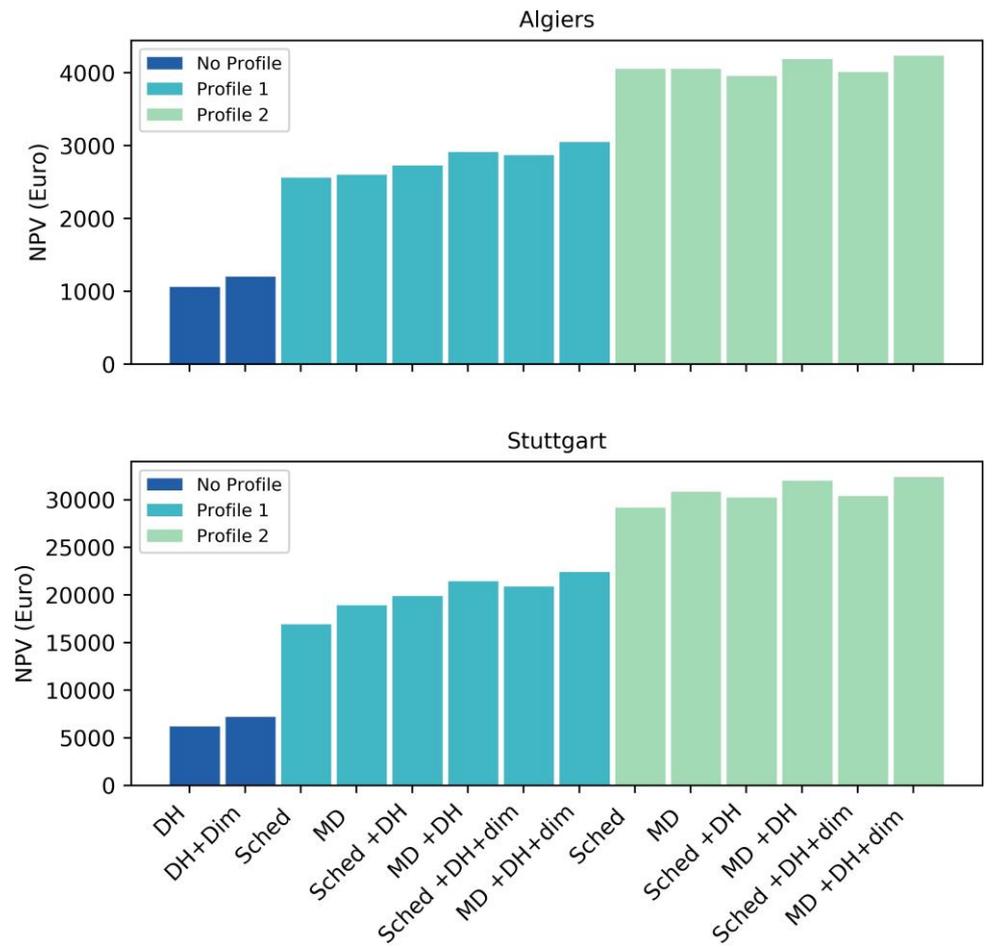

Figure 5: Net present value (NPV) for lighting control strategies

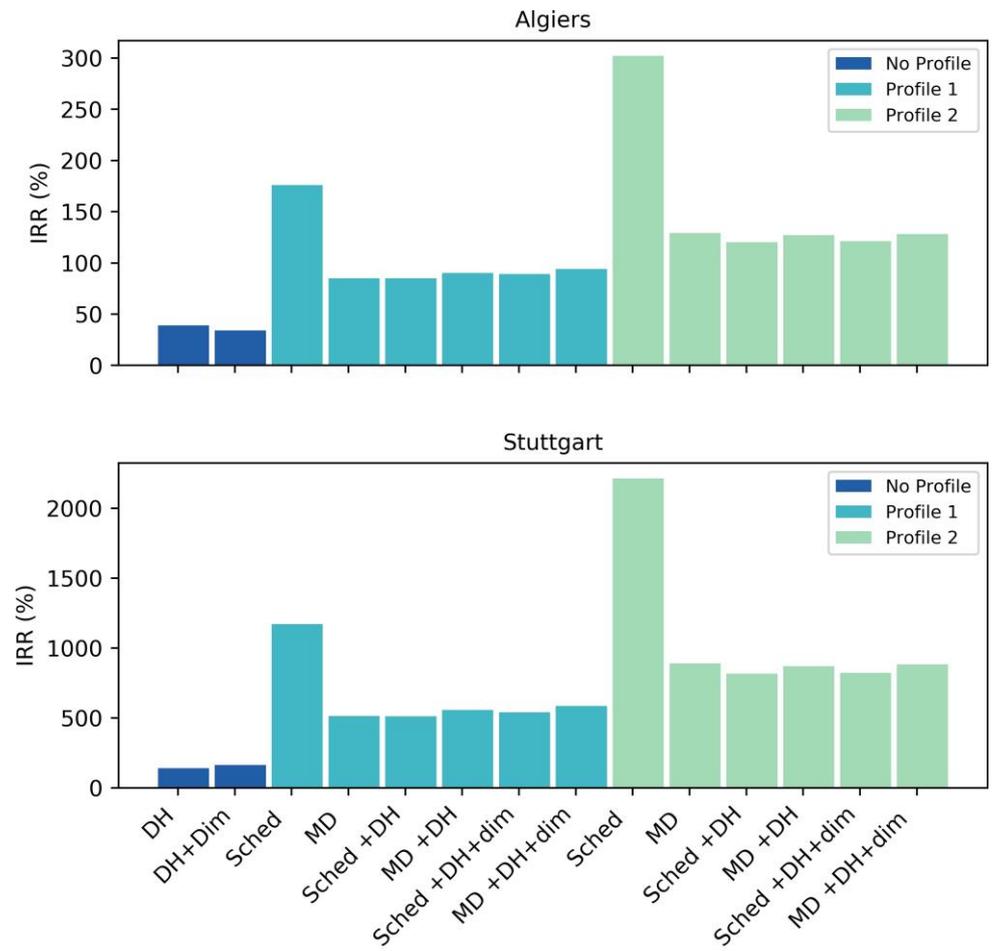

Figure 6: Internal rate of return (IRR) for lighting control strategies

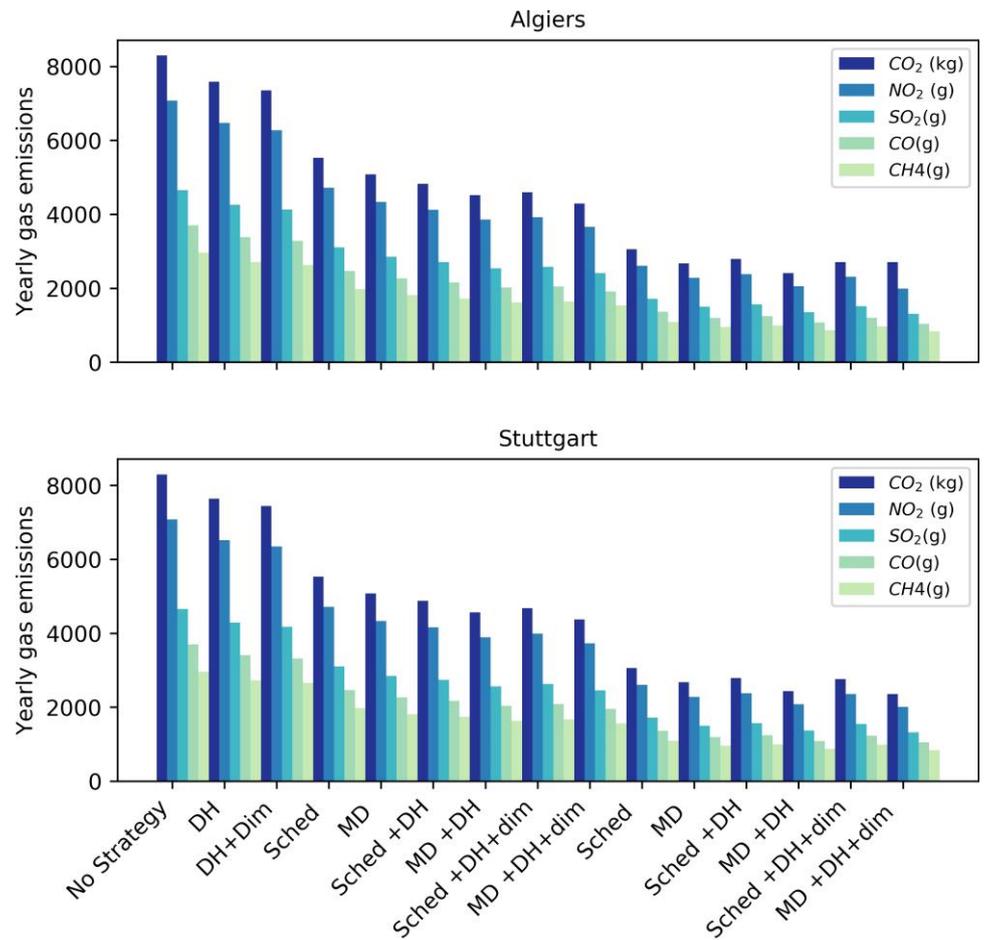

Figure 7: Gas emissions for lighting control strategies

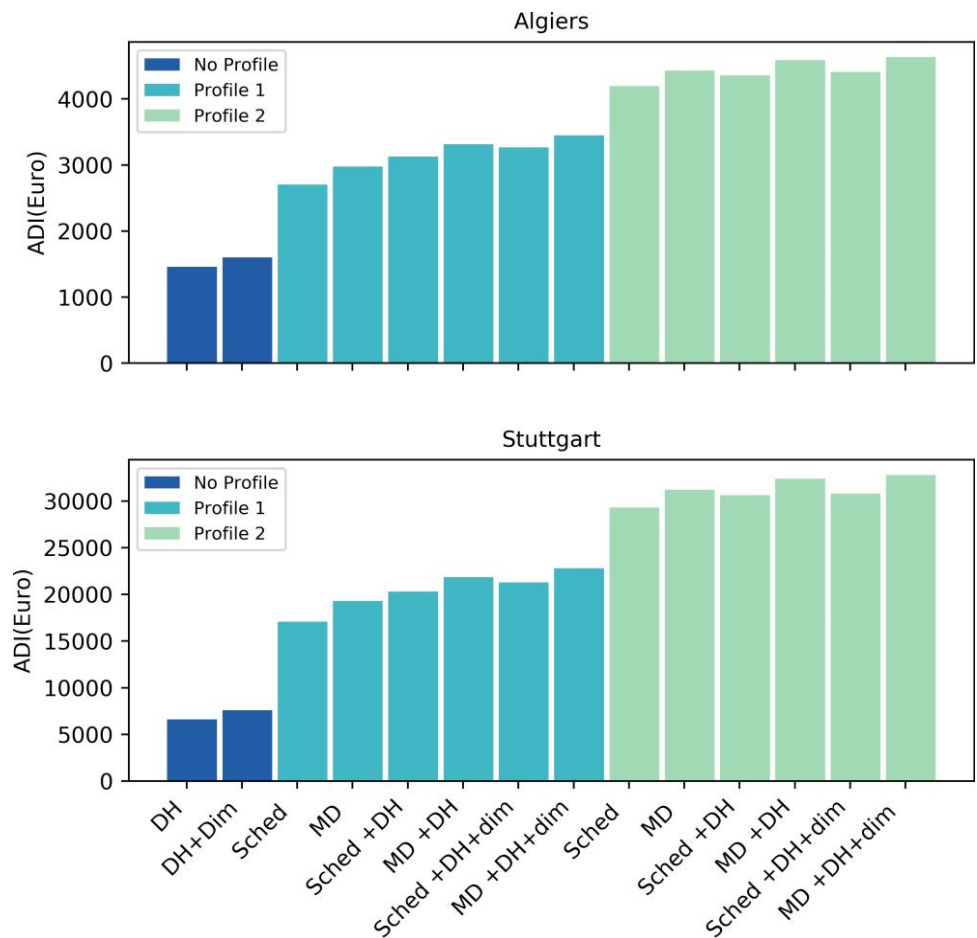

Figure 8: Social Metric: Additional disposal income for lighting control strategies

## 6. Conclusion

Buildings Energy management technologies have the technical potential to provide energy saving benefits for both users and utilities. These products deliver demand reductions by using contextual information such as occupancy and available daylight and can support users in reducing or shifting loads by sending notifications and displaying energy usage. The currently proposed products on the market incorporate both energy and non-energy related smart home devices. This combination presents the value of smart home technology for consumers, enabling home energy management while increasing security and comfort, e.g., being able to reduce light load while controlling it vocally.

This research provided insights into the energy management gain from each one of the most common energy-related features for light control. Contrary to the existing studies, our study is the first to investigate into a more fine-grained analysis of technology choices. We enlarge the analysis which classically focuses on energy reduction to economic, environmental and social considerations. Separating features allows to separate energy and non-energy functionality and shows how much they are contributing to energy saving. The result has shown that smart home technologies deliver a variety of benefits, but depending on users profiles, energy cost, and the geographical region, the amount of saved energy for each feature changes. As an example, dimming and daylight harvesting are less useful for a family whose members are absent during daytime, and scheduling can be an efficient, low-cost solution for a family with regular occupancy and vacancy periods. Whereas producers and governments strongly hightlight the potential benefit in terms of energy savings, our study shows that including economic and environmental concerns can act as a strong motivator to engage into building automation.

Because the most impacting factor is the time of operation, which is highly related to the user (comparing Profile 1 and Profile 2), it is beneficial to include in commercialized devices the ability to monitor and track the users' behavior accurately, and then accordingly optimize actuation. Coupling sensing devices with artificial intelligence (AI) improves the accuracy of detection and leads to increased energy savings. While state-of-the-art products provides such solutions[46], they are less present in commercialized lightingdevices compared to other smart home systems such as thermostats, security cameras, wireless speakers, etc. Shortly, modern solutions expected to contribute in changing behaviors towards ecological friendly habits through incentive services based on advanced technologies such artificial intelligence (AI), IoT, wireless communications and networking, which will contribute to shifting the current building automation towards digital and software-defined buildings.

## Acknowledgements

This work is supported with a grant from the Arab German Young Academy (AGYA), under the smart city collaborative project at its innovation working group project and a tandem project. It draws on financial support of the German Federal Ministry of Education and Research (BMBF) grant 01DL16002. The work has been carried out in part at CERIST research center Algeria, which is supported by the Algerian Ministry of Higher Education through the DGRSDT, and in part at HfWU Nrtingen, Germany.